\documentclass[fleqn,preprint,5p]{elsarticle}

\usepackage{chngcntr}

\usepackage{graphicx}
\usepackage{bm}
\usepackage{dcolumn}
\usepackage{color}

\usepackage{graphics}
\usepackage[retainorgcmds]{IEEEtrantools}
\usepackage{xspace} 
\usepackage{slashed}
\usepackage{amsmath}
\usepackage{amsfonts}
\usepackage{amssymb}

\newcommand{\fixme}[1]{}
\newcommand{\cb}[1]{}
\newcommand{\mc}[1]{}

\newcommand{\indmode}{\ell\omega} 
\newcommand{\indmodem}{\ell,-\omega} 
\newcommand{\Glw}[1]{G_{\indmode}^{#1}}
\newcommand{\Rin}{R^{\text{in}}_{\indmode}}
\newcommand{\Rup}{R^{\text{up}}_{\indmode}}
\newcommand{\Rinm}{R^{\text{in}}_{\indmodem}}
\newcommand{\Rupm}{R^{\text{up}}_{\indmodem}}
\newcommand{\Rinup}{R^{\text{in/up}}_{\indmode}}
\newcommand{\Rincc}{R^{\text{in}^*}_{\indmode}}
\newcommand{\Rupcc}{R^{\text{up}^*}_{\indmode}}
\newcommand{\Rinref}{R^{\text{in,ref}}_{\indmode}}

\newcommand{\Rintra}{R^{\text{in,tra}}_{\indmode}}
\newcommand{\Rintram}{R^{\text{in,tra}}_{\indmodem}}
\newcommand{\Rupref}{R^{\text{up,ref}}_{\indmode}}

\newcommand{\Ruptra}{R^{\text{up,tra}}_{\indmode}}
\newcommand{\Ruptram}{R^{\text{up,tra}}_{\indmodem}}

\begin{document}

\counterwithout{equation}{section}

\author{Claudia Buss}
\address{Aix Marseille Univ, CNRS, LAM, Laboratoire d'Astrophysique de Marseille, Marseille, France.}
\address{Centro Brasileiro de Pesquisas F\'isicas (CBPF), Rio de Janeiro, CEP 22290-180, Brazil.}

\ead{claudia.buss@lam.fr.}


\author{Marc Casals}
\address{Centro Brasileiro de Pesquisas F\'isicas (CBPF), Rio de Janeiro, CEP 22290-180, Brazil.}
\address{School of Mathematics and Statistics and   UCD Institute for Discovery, University College Dublin, Belfield, Dublin 4, Ireland.}
\ead{mcasals@cbpf.br, marc.casals@ucd.ie.}


\title{Quantum correlator outside a Schwarzschild black hole}

\begin{abstract} 
We 
calculate the quantum correlator in Schwarzschild black hole space-time.
We perform the calculation for a scalar field in three different quantum states: 
 Boulware,  Unruh  and  Hartle-Hawking,
and for points along a timelike circular geodesic.
The results show that the correlator presents a global fourfold singularity structure, which is state-independent.
Our results
 also show the different correlations in the three different quantum states arising in-between the singularities.
\end{abstract}

\date{\today}
\maketitle

\section{Introduction}
\mc{
Things to do:
\begin{enumerate}
\item include the analytical value of the integrand at $\omega=0$ - this should fix the wrong behaviour at $t \gtrsim 78$ (a rather cheap alternative would be to cut off the plots from that time onwards)
\item can we get any physics from this calculation? is there a similar calculation of FGF for points separated in a different way (though note that points must be outside a normal nbd so that the calculation is `easily doable') from which we could get any physics?
\item I don't know if we could get much physics out of the following one, but it should be quite easy to do and it'd be interesting/curious/new to do a calculation of FGF for points spacelike-separated: pick a value of $d\varphi/d\tau$ large enough so that the corresponding circular (still at $r=6M)$ worldline (whether geodesic or not) where $x$ and $x'$ lie is spacelike
\end{enumerate}
}

The Feynman Green function (FGF) for a quantum `matter' field propagating on a classical, curved background space-time is important for various reasons.
One of the reasons is that one may obtain the (expectation value of the) quantum stress-energy tensor by applying a certain operator on the FGF.
In its turn, the quantum stress-energy tensor  is a crucial quantity within semiclassical gravity:
when appropriately renormalized,
 it replaces the classical stress-energy tensor in the classical Einstein equations.
Solving the semiclassical equations provides the  backreaction due to the quantum matter field on the classical background space-time on which it propagates
(see, e.g.,~\cite{Birrell:Davies}).

In this paper we are interested in the FGF for the following different reason. The FGF is a function of two space-time points and it
provides the quantum correlations 
 between these two points.
In the case of a Schwarzschild black hole space-time, for example, one would expect to see correlations between quantum Hawking `particles', which escape to infinity, and their counterparts, which fall into the 
black hole~\cite{hawking1975particle}. 
Whereas Hawking radiation is too weak to be detected in an astrophysical setting, an analogue of the correlations 
between Hawking particles
has recently been observed in condensate systems set up as analogue black holes models~\cite{Steinhauer:2015saa}.

In the calculation of the quantum stress-energy tensor, one must in principle take the coincidence limit of the two space-time points in the FGF.
It is well-known, however, that the FGF diverges in  this limit. Therefore, one must perform an appropriate renormalization so as to obtain a renormalized  quantum stress-energy tensor that is to be inserted in the semiclassical Einstein equations.
In the case that interests us here, on the other hand, the points are kept separated and so we are spared the arduous task of renormalization.
However, the FGF  does not only diverge at coincidence but also whenever the two space-time points are connected by a null geodesic (e.g.,~\cite{Garabedian,Ikawa}).
These divergences are `physical', they are not to be renormalized away, and so one must 
embrace
 them.
 Mathematically, they are linked to the fact that the FGF is a bi-{\it distribution}.
As a consequence, the calculation of the FGF in Schwarzschild space-time is a highly non-trivial task also when the points are separated.

In this paper, we calculate the quantum correlator, FGF, for a massless scalar field on a Schwarzschild black hole space-time.
Our calculation is semi-analytical and is  for points outside the horizon -- specifically,  along a timelike circular geodesic.
We obtain the FGF when the field is in three different  quantum states of physical interest: the Boulware state~\cite{boulware1975quantum,boulware1975spin} (representing
a cold star), the Unruh state~\cite{unruh1976notes} (representing an evaporating black hole)
and the Hartle-Hawking state~\cite{hartle1976path} (representing a black hole in thermal equilibrium).

To the best of our knowledge, this is the first time that the quantum correlator has been explicitly calculated in Schwarzschild space-time
for separated points.
The separation of the points along a timelike circular geodesic allows us to observe the `physical' divergences of the FGF mentioned above.
This calculation manifests a fourfold singularity structure in the FGF as the null wavefront passes through caustic points
(points where neighbouring null geodesics focus) of the background Schwarzschild space-time.
The real part of the FGF is essentially the retarded Green function (RGF).
We use that as a check of our results: we verify  that  the real part of our calculation of the FGF agrees with existing literature results for the RGF~\cite{CDOW13}, for which the fourfold structure is already known~\cite{Ori1short,CDOWa,Zenginoglu:2012xe,casals2016global,Dolan:2011fh,harte2012caustics}.
All the information about the quantum state of the field, however, is contained in the {\it imaginary} part of the FGF, which is not obtainable in terms of the RGF.
We find a fourfold singularity structure in the imaginary part of the FGF which is analogous to that in the real  part of the FGF (or, equivalently, in the RGF).
Specifically, we find that the structure in the imaginary part of the FGF is
\begin{equation}
\text{PV}\!\left(\!\frac{1}{\sigma}\!\right)\to-\delta(\sigma)\to-\text{PV}\!\left(\!\frac{1}{\sigma}\!\right)\to\delta(\sigma)\to \text{PV}\!\left(\!\frac{1}{\sigma}\!\right)\to \cdots
\label{4-fold FGF intro}
\end{equation}
where $\delta$ is the Dirac-delta distribution and PV denotes the Cauchy principal value distribution. 
Here,  $\sigma$ is  Synge's world-function (i.e., one-half of the squared distance along the -unique- geodesic connecting the two space-time points), but appropriately extended to be valid globally (see~\cite{casals2016global}).
The structure in Eq.\eqref{4-fold FGF intro}  is as in
the already known  structure in the real part of the FGF, and so in the RGF, but shifted by one fold.
This singularity structure of the imaginary part of the FGF that our semi-analytic results show   had been conjectured in~\cite{CDOWa,Casals:2012px,casals2016global} but, to the best of our knowledge, had not been shown before.
This structure is independent of the quantum state, and so the different correlations in the different quantum states lie in-between these singularities, which 
our results also show.

The layout of the rest of this paper is as follows.
In Sec.\ref{sec:QFT} we  give the expressions for the FGF in the different quantum states.
In Sec.\ref{sec:sing} we discuss the global singularity structure of the FGF.
In Sec.\ref{sec:method} we describe the method used to evaluate the expressions for the FGF.
We present the results of the evaluation in Sec.\ref{sec:results}.
We conclude in Sec.\ref{sec:end}.
We choose units $c = G =\hbar= 1$ and metric signature $(-+++)$ \mc{and $\hbar$?}.

\section{Quantum Correlator on Schwarzschild Space-time}\label{sec:QFT}
 
We consider a masless scalar field propagating on \\ Schwarzschild space-time.
The corresponding FGF, $G_F(x,x')$, depends on two space-time points: a base point $x$ and a field point  $x'$.
It
satisfies the Klein-Gordon wave equation with a $4$-dimensional invariant Dirac distribution as the source:
\begin{equation}\label{eq:FDF eq}
\Box G_F(x,x')=-\frac{\delta^{(4)}(x-x')}{\sqrt{|g|}},
\end{equation}
where $\Box$ is the D'Alembertian in Schwarzschild space-time and $g=-r^4\sin^2\theta$ is the determinant of the  metric in Schwarzschild co-ordinates.
In these co-ordinates, the space-time points are given by
$x=\{t,r,\theta,\varphi\}$ and
$x'=\{t',r',\theta',\varphi'\}$.
Without loss of generality, we set $t'=0$.

The Klein-Gordon equation in Schwarzschild space-time separates in Schwarzschild co-ordinates and so
its solution admits a straight-forward mode decomposition. 
When $t>0$, as 
we henceforth take it to be the case,
 the FGF is given by~\cite{Candelas:1980zt}
\begin{align}\label{eq:GF}
&
G_F^{\Psi}(x,x')=
\nonumber \\ &
\frac{i}{\left(4\pi\right)^2}\sum_{\ell=0}^{\infty}(2\ell+1)P_{\ell}(\cos\gamma)\int_{-\infty}^{\infty}\frac{d\omega}{\omega}
\Glw{\Psi}(r,r';t),
\end{align}
where
$\gamma$ is the angular separation between the two points and
 $\Glw{\Psi}=\Glw{\Psi}(r,r';t)$ 
are  modes whose expression depends on the quantum state $\Psi$ of the field.

In Schwarzschild space-time there are three quantum states of interest.
The Boulware state~\cite{boulware1975quantum,boulware1975spin} is irregular on  both the future and past horizons and is empty at radial infinity; it
it is thus said to represent a cold star.
The Unruh state~\cite{unruh1976notes} is regular on the future horizon, irregular on the past horizon, empty at past null infinity and contains Hawking radiation
going out to future null infinity; it thus represents an astrophysical black hole evaporating via emission of Hawking radiation.
Finally, the Hartle-Hawking state~\cite{hartle1976path} is regular everywhere and it represents a black hole in (unstable) thermal equilibrium with its
own quantum radiation~\cite{kay1991theorems,sanders2015construction}.
The temperature of the Hawking radiation is $T=\kappa/(2\pi)$, where $\kappa=1/(4M)$ is the surface gravity of the black hole of mass $M$.

When the field is in the Boulware ($\Psi=B$), Unruh ($\Psi=U$) and Hartle-Hawking ($\Psi=H$) state, the modes $\Glw{\Psi}$ are respectively given by~\cite{Candelas:1980zt}
\begin{equation}\label{eq:FGF modes B}
\Glw{B}=
e^{-i\omega t}\theta(\omega)\left(\Rup(r)\Rupcc(r')+\Rin(r)\Rincc(r') \right),
\end{equation}

\begin{equation}\label{eq:FGF modes U}
\Glw{U}=
e^{-i\omega t}\left(
\frac{
\Rup(r)\Rupcc(r')
}
{1-e^{-2\pi\omega/\kappa}}
+\theta(\omega)\Rin(r)\Rincc(r') \right),
\end{equation}
and
\begin{equation}\label{eq:FGF modes H}
\Glw{H}=
e^{-i\omega t}\frac{\Rup(r)\Rupcc(r')}{1-e^{-2\pi\omega/\kappa}}
+e^{i\omega t}\frac{\Rin(r)\Rincc(r') }
{e^{2\pi\omega/\kappa}-1}.
\end{equation}

The radial modes $\Rinup$ are solutions of the homogeneous radial equation:
\begin{equation} \label{eq:radial ODE}
\left(\frac{d^2}{dr_*^2}+\omega^2-V(r)\right)R(r)=0,
\end{equation}
\begin{equation}\label{eq:potential}
V(r)\equiv 
\left(1-\frac{2M}{r}\right)\left(\frac{\ell(\ell+1)}{r^2}+\frac{2M}{r^3}\right),
\end{equation}
where $r_*\equiv r+2M\ln\left|\frac{r}{2M}-1\right|\in(-\infty,\infty)$,
 obeying certain, ingoing/upgoing boundary conditions.
These conditions are:
 \begin{align}
\label{eq: bc Rin}
\Rin\sim
\begin{cases}
 \Rintra e^{-i\omega r_*},&\!\!\!  r_*\to -\infty,\\
 e^{-i\omega r_*}+\Rinref e^{+i\omega r_*},&\!\!\!  r_*\to +\infty,
\end{cases} 
\end{align}
and 
\begin{align}
\label{eq: bc Rup}
\Rup\sim
\begin{cases}
 e^{+i\omega r_*}+ \Rupref e^{-i\omega r_*},&\!\!\!  r_*\to -\infty,\\
\Ruptra e^{+i\omega r_*},&\!\!\!  r_*\to +\infty,
\end{cases} 
\end{align}
where
$R_{\indmode}^{\text{in,ref/tra}}$ are the reflection/transmission
 coefficients of the ingoing solution; similarly  
 $R_{\indmode}^{\text{up,ref/tra}}$
  for the upgoing solution.
  
  In this paper we present results of the explicit evaluation of Eq.\eqref{eq:GF} for $\Psi=$ B, U and H.
  Before that, however, we discuss the global singularity structure of the FGF.


\section{Conjectured Singularity Structure}\label{sec:sing}

\begin{figure}
\centering
\includegraphics[width=8.5cm]{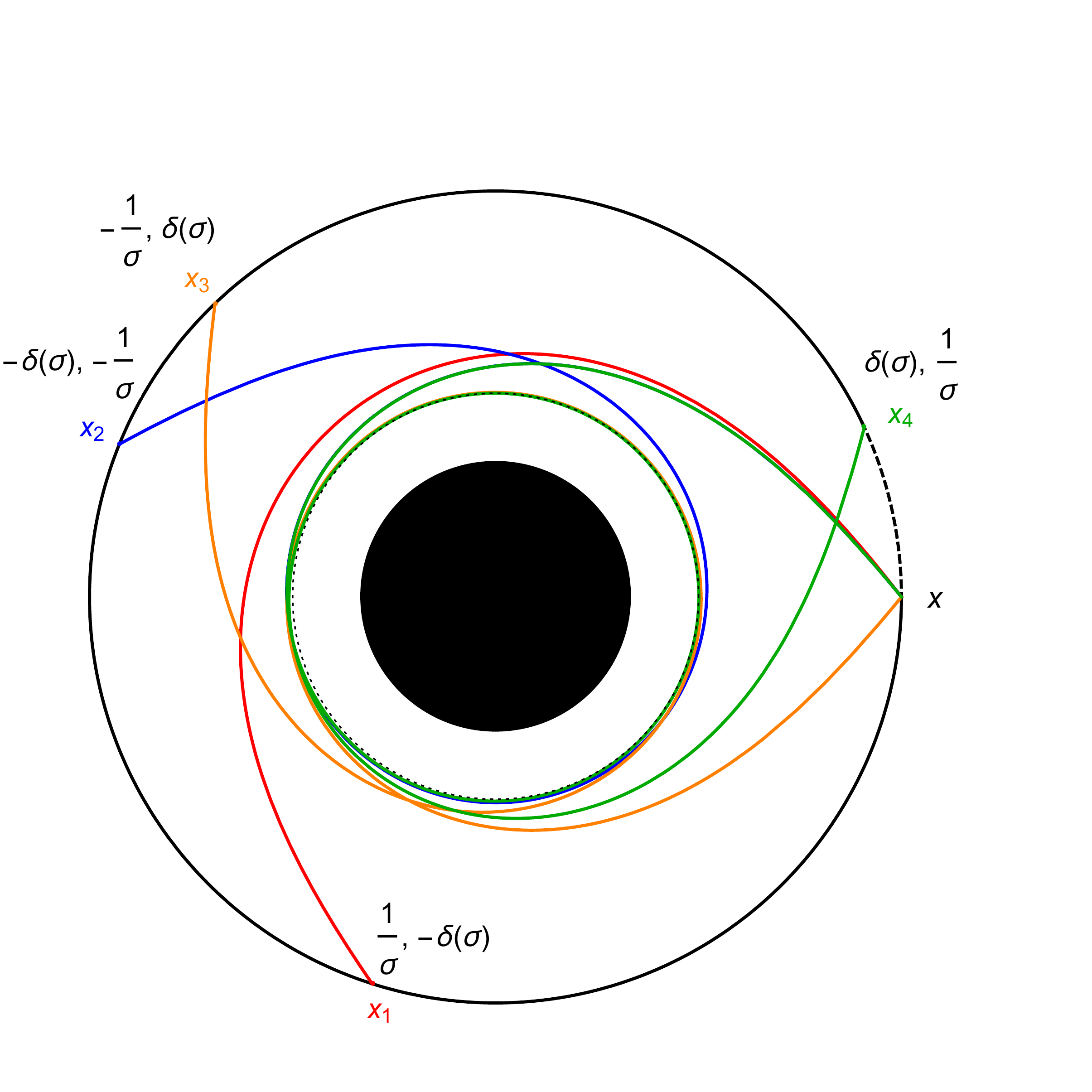}
\caption{
Plot of  geodesics on a Schwarzschild black hole space-time together with the global singularity structure
of: 
real, imaginary parts of the 
Feynman Green function (as per Eqs.\eqref{4-fold} and \eqref{4-fold FGF}, respectively).
Black curve: timelike circular geodesic at $r=6M$.
Coloured curves: null geodesics which emanate from an arbitrary space-time point ($x$) on the timelike geodesic and meet
another point ($x_i$, $i=1,2,3,4$) on the timelike geodesic at a different time.
}
\label{fig:geods}
\end{figure}

The so-called Hadamard form~\cite{Hadamard} for a Green function is an analytic expression which makes explicit the divergence of the Green function when the two space-time points are connected by a null geodesic.
The Hadamard form, however, has the drawback that it is only valid 
locally. Specifically, it is only valid within a normal neighbourhood $\mathcal{N}(x)$ of the base point $x$~\cite{Friedlander}:
a neighbourhood such that every point $x'\in \mathcal{N}(x)$  is connected to $x$ by a unique geodesic which lies in $\mathcal{N}(x)$.
For example, 
consider a timelike circular geodesic at $r=6M$ in Schwarzschild space-time, as represented in Fig.\ref{fig:geods},
and  an arbitrary point $x$ on it.
Then, a discrete number of points $x_1,x_2, x_3\dots$ on that geodesic are connected to $x$, not only  by  that timelike geodesic, but also by a null geodesic;
we say that $x_1,x_2, x_3\dots$ are light-crossings.
Thus, the first light-crossing
$x_1$ separates points on the circular geodesic which  lie in $\mathcal{N}(x)$ from points (including $x_1$)  which do not.

The Hadamard form, which is valid $\forall x'\in\mathcal{N}(x)$,  for the FGF is (e.g., \cite{DeWitt:1960})
\begin{equation}\label{eq:Feynman GF Had}
G_F(x,x')=\lim_{\epsilon \rightarrow 0^+}
\frac{i}{2\pi}
\left[\frac{U}{\sigma+i\epsilon}-V\ln\left(\sigma+i\epsilon\right)+W\right],
\end{equation}
where  $U=U(x,x')$, $V=V(x,x')$ and $W=W(x,x')$ are regular and real-valued biscalars and $\sigma=\sigma(x,x')$ is the so-called Synge's world function.
This function is equal to one-half of the square of the  geodesic distance joining $x$ and $x'$.
We note that while $U$ and $V$ are determined {\it uniquely} by the geometry of the space-time, $W$ is not;
the value of $W$ is in principle different for different quantum states.

The retarded Green function (RGF), $G_R(x, x')$, satisfies the Green function equation \eqref{eq:FDF eq}
 with the boundary condition that it is zero if $x'$ is not in the causal future of $x$.
The RGF is related to the FGF (in any quantum state $\Psi$) via (e.g.,~\cite{Birrell:Davies,DeWitt:1960})
\begin{equation}\label{eq:RGF via FGF}
G_R(x,x')=
2\theta(t)\text{Re}\left(G_F^{\Psi}(x,x')\right).
\end{equation} 
Eqs.(\ref{eq:Feynman GF Had}) and (\ref{eq:RGF via FGF}),
together with the
 distributional properties 
\begin{equation}\label{eq:PV}
\lim_{\epsilon \rightarrow 0^+}\frac{1}{\sigma + i \epsilon} =\text{PV}\left({\frac{1}{\sigma}} \right)-i\pi \delta(\sigma),
\end{equation}
and
\begin{equation}\label{eq:ln}
\lim_{\epsilon \rightarrow 0^+} \ln\left(\sigma+i\epsilon\right)=\ln|\sigma|+i\pi\theta(-\sigma),
\end{equation} 
 imply
that the Hadamard form  for the RGF is given by
\begin{equation}\label{eq:Hadamard RGF}
G_R(x, x') =\left(U(x,x')\delta(\sigma)+ V(x,x')\theta(-\sigma)\right)\theta_+(x,x').
\end{equation}
Here, $\theta_+(x, x')$ equals $1$ if $x'$ lies to the future of $x$ and equals $0$ otherwise.

As mentioned, the Hadamard form is in principle not valid when
$x'\notin \mathcal{N}(x)$,
which is generally the case 
when the points are `far enough' in a curved black hole space-time such as Schwarzschild.
Despite that, it is known~\cite{Garabedian,Ikawa} that a Green function  diverges when the two space-time points are connected via  a null geodesic, no matter how `far' the two
points are. The form of these global singularities in the case of a black hole space-time, however, was not known until recently.
In a series of papers, the divergence of the RGF for a{\it rbitrary} null-separated points in Schwarzschild space-time (as well as other space-times, such as Kerr and black hole toy models) has been obtained in~\cite{casals2016global,Ori1short,CDOWa,Dolan:2011fh,harte2012caustics,Casals:2012px,Zenginoglu:2012xe,Yang:2013shb}. 
These papers show that the divergence of $G_R$ follows a fourfold pattern.
Specifically, the pattern for the leading divergence in $G_R$ is
\begin{equation}
\delta(\sigma)\to\text{PV}\left(\frac{1}{\sigma}\right)\to-\delta(\sigma)\to-\text{PV}\left(\frac{1}{\sigma}\right)\to\delta(\sigma)\to\cdots
\label{4-fold}
\end{equation}
and for the sub-leading divergence it is
\begin{equation}
\theta(-\sigma) \to -\ln\left|\sigma\right| \to -\theta(-\sigma) \to \ln\left|\sigma\right| \to \theta(-\sigma)\to\cdots
\label{4-fold,sublead}
\end{equation}
The singularity type changes as the null wavefront passes through caustic points (which
have $\gamma=0$ or $\pi$).
In~\cite{casals2016global,Zenginoglu:2012xe} it was shown that an exception to the above fourfold structure is
at caustic points,
 where the structure is twofold instead.

In the particular case of Fig.\ref{fig:geods}, 
this means that
$G_R$ will diverge at the light-crossings $x_1,x_2, x_3\dots$  and that
 its leading singularity 
 at $x_1$ will be 
 `$\text{PV}(1/\sigma)$', at $x_2$  it will be  `$-\delta(\sigma)$', at $x_3$ it will be  `$-\text{PV}(1/\sigma)$' and at $x_4$ it will be `$\delta(\sigma)$';
its subleading singularity will respectively be 
`$-\ln\left|\sigma\right|$', `$-\theta(-\sigma)$', `$\ln\left|\sigma\right|$' and `$\theta(-\sigma)$'.
Note that, in this case, the singularity `$\delta(\sigma)$' and discontinuity `$\theta(-\sigma)$' in the Hadamard form would take place at coincidence, $x=x'$.

We note that the world function $\sigma=\sigma(x,x')$ is strictly well-defined only
for $x'\in\mathcal{N}(x)$.
This is because  if there is more than one geodesic (lying in $\mathcal{N}(x)$) joining $x'$ and $x$, then $\sigma$ is no longer uniquely defined.
However, by indicating along which geodesic $\sigma$ is calculated, this biscalar can effectively be extended to any pairs of points in Schwarzschild space-time -- see~\cite{casals2016global} for details. 
It is in this extended sense that we are using $\sigma$ outside the region of validity of the Hadamard form.

As opposed to the RGF (which is directly related to the real part of the FGF via Eq.\eqref{eq:RGF via FGF}), to the best of our knowledge, 
the {\it global} singularity structure of the imaginary part of the FGF in Schwarzschild space-time has not yet been obtained.
In~\cite{CDOWa,Casals:2012px,casals2016global} it was conjectured that the structure for the imaginary part of the FGF would be the following (except at caustics):
\begin{equation}
\text{PV}\!\left(\!\frac{1}{\sigma}\!\right)\to-\delta(\sigma)\to-\text{PV}\!\left(\!\frac{1}{\sigma}\!\right)\to\delta(\sigma)\to \text{PV}\!\left(\!\frac{1}{\sigma}\!\right)\to \cdots
\label{4-fold FGF}
\end{equation}
We note that this is like the fourfold singularity structure in Eq.\eqref{4-fold} for the RGF but shifted by one fold.

The Hadamard form Eq.\eqref{eq:Feynman GF Had}, together with Eq.\eqref{eq:PV}, only provides the first term in Eq.\eqref{4-fold FGF};
the rest of the terms were a conjecture.
This conjecture 
was based on tentatively allowing for the form in Eq.\eqref{eq:Feynman GF Had} to be essentialy valid (although with the mentioned appropriate extension of $\sigma$)
outside a normal neighbourhood.
By using the fact that $U(x,x')$ obeys a transport equation along a geodesic, it can be argued~\cite{CDOWa} that it
 picks up a phase of `$-\pi/2$' as the geodesic crosses a caustic point (see, e.g.,~\cite{B&M} for a similar phenomenon outside General Relativity).
  That is, a factor of `$-i$' would be picked up by $U$ at every caustic, which, combined with  Eq.\eqref{eq:PV} and the first term in Eq.\eqref{eq:Feynman GF Had}, 
  would yield Eq.\eqref{4-fold FGF}.
We note that the exact value of the phase picked up by $U$ would affect the singularity cycle and that varies with the space-time.
For example, from specific calculations, it seems that space-times with caustics for which the Hadamard tail $V(x,x')$ is non-zero
possess a similar four-fold pattern  (apart from Schwarzschild as shown here for RGF and FGF, it has been observed for the RGF in Kerr~\cite{harte2012caustics}, Nariai~\cite{CDOWa} and 
Pleba\'nski-Hacyan~\cite{Casals:2012px} space-times), whereas   space-times with caustics for which $V(x,x')\equiv 0$ possess instead a two-fold pattern (such is  the case of the Einstein Static Universe~\cite{BGO} and Bertotti-Robinson space-time~\cite{ottewill2012quantum}).

 Fig.\ref{fig:geods} indicates the leading-order divergences in Eqs.\eqref{4-fold} and \eqref{4-fold FGF} for, respectively, the real and imaginary parts of the FGF (in the real part case, it
 is of course equivalent to the structure of the RGF),
for the case of the timelike circular geodesic.
The results of the semi-analytic
 calculation that we present in the next section 
show that 
the conjecture in Eq.\eqref{4-fold FGF}
 is correct (at least for the case that we calculated the FGF, i.e., for the timelike circular geodesic).

\section{Method}\label{sec:method}

In this section we  describe the method we used to calculate the FGF.
Using  Eq.\eqref{eq:GF} to calculate the FGF is 
a challenging
task, particularly since the FGF is a bi-distribution which diverges not only at coincidence but also at  light-crossings, as described in the previous section.
Technically, Eq.\eqref{eq:GF} involves both an integral and an infinite sum, which does not converge at  light-crossings.
A similar mode-sum calculation of a Green function in Schwarzschild space-time was successfully achieved in~\cite{CDOWa}.
The difference is that the mode-sum calculation in~\cite{CDOWa} was of the RGF and achieved by deforming the integral on the complex-frequency plane, whereas 
the calculation here is of the FGF and we achieve it by integrating directly over real frequencies.
Similar calculations by integrating  over real frequencies but for the RGF instead of FGF have been achieved in~\cite{WardellRealFreqRGFSchw,GFKerr}.

We first note that, in the practical calculation, we  `folded up' the integrals in Eq.\eqref{eq:GF} over $\omega: -\infty\to\infty$ so that they instead run over $\omega: 0\to\infty$.
We achieve this by using the symmetries $\Rinm/\Rintram=\left(\Rin/\Rintra\right)^*$
and $\Rupm/\Ruptram=\left(\Rup/\Ruptra\right)^*$, which are valid for all $\omega\in \mathbb{R}$.
Once the integrals have been folded up to run over $\omega: 0\to\infty$, one can explicitly show~\cite{Buss-2016} that the real parts of the corresponding integrands of the correlators  
for the Unruh, Boulware and Hartle-Hawking states are equal.
In this sense, the equalities $\text{Re}\left(G_F^{B}\right)=\text{Re}\left(G_F^{U}\right)=\text{Re}\left(G_F^{H}\right)$ (as it should be, from Eq.\eqref{eq:RGF via FGF})
are satisfied {\it mode-by-mode}.

In practise, one must implement some cutoffs  $\ell_{max}$ in the $\ell$-sum and $\omega_{max}$ in the $\omega$-integral in Eq.\eqref{eq:GF}.
Because of these cutoffs, not only the divergences of the FGF are `smoothed out', but also, if one sums and integrates the modes directly as in Eq.\eqref{eq:GF},
spurious oscillations appear. We thus followed~\cite{CDOWa,WardellRealFreqRGFSchw} and multiplied the modes by   `smoothing' factors (for further justification, see~\cite{Hardy}).
Specifically, we found it convenient to introduce the smoothing factor
`$\text{Exp}\left(-\ell^2/(2\ell_{cut}^2)\right)$' for the $\ell$-sum and
`$\left(1-\text{Erf}\left(2M(\omega-\omega_0)\right)\right)/2$'   for the $\omega$-integral, where `$\text{Erf}$' is the error function and $\ell_{cut},\omega_0>0$ are  parameters.
We found that the following choices of values worked well: $\ell_{max}=100$ and $M\omega_{max}=10$ as cutoff parameters; $\ell_{cut}=12$
and $M\omega_0=8.5$ as smoothing parameters. Also, we calculated the modes at discrete $\omega$-values using  a 
stepsize of  
$\Delta \omega=10^{-3}/M$. 
We note that
 increasing $\ell_{cut}$ would `sharpen' the divergences at light-crossings but, on the other hand, it would allow for more pronounced spurious
oscillations near the divergences.
In its turn, a smaller value for $\Delta \omega$ leads to a finer grid near $\omega=0$ and so to more accurate results at later times -- 
as an example, we found that, in our case, taking $\Delta \omega=10^{-2}/M$ instead of $\Delta \omega=10^{-3}/M$ leads to visually-wrong results for $t$ larger than
about $65M$.

The modes $\Glw{\Psi}$ depend on the radial solutions $\Rinup$.
In order to calculate them, we used the
 semi-analytical method
 of   Mano, Suzuki and Takasugi (MST; see~\cite{Sasaki:2003xr} for a review and~\cite{Casals:Ottewill:2015,casals2016high} for further details and extension of the method).
Essentially, the MST method  consists of finding the radial solutions $\Rinup$ and their radial coefficients via infinite series of special functions (such as hypergeometric and confluent
hypergeometric functions).
After calculating the FGF modes (including the mentioned smoothing factors) 
for the Boulware, Unruh and Hartle-Hawking states,
at the indicated $\ell$ values and discrete frequencies, we interpolated the $\omega$-integrands
and integrated them using the software {\rm MATHEMATICA}.
Further details --although applied to the calculation of the RGF-- will be provided in~\cite{GFKerr}.
With this data we  constructed the FGF using Eqs.\eqref{eq:FGF modes B}--\eqref{eq:FGF modes H}.
We also used this radial data to construct the RGF using Eqs.(2.34) and (2.35) in~\cite{Casals:Ottewill:2015}.
In the next section  we present the results obtained.

\section{Results for the Quantum Correlator}\label{sec:results}

\mc{Do zoomed-in versions of the plots; do non-log versions of the plots to show 4-fold structure}

We applied the method described in the previous section to the calculation of the scalar FGF in Schwarzschild space-time
for points along a timelike circular geodesic at $r=6M$, which is represented in Fig.\ref{fig:geods}.
For comparison purposes, we also calculated the RGF using the method of complex-frequency integration of~\cite{CDOWa}.

\begin{figure}
\centering
\includegraphics[width=8.5cm]{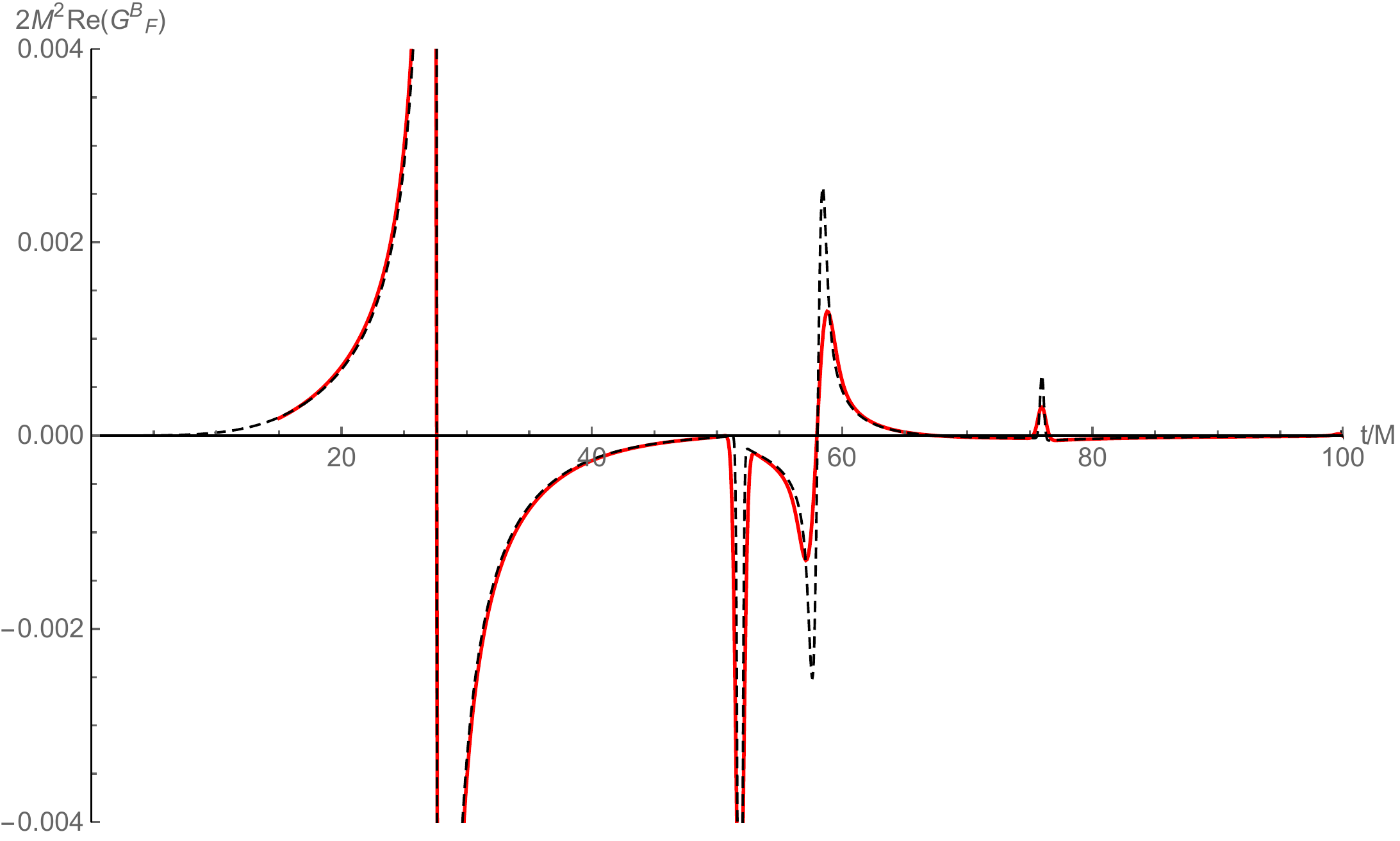}
\\
 \includegraphics[width=8.7cm]{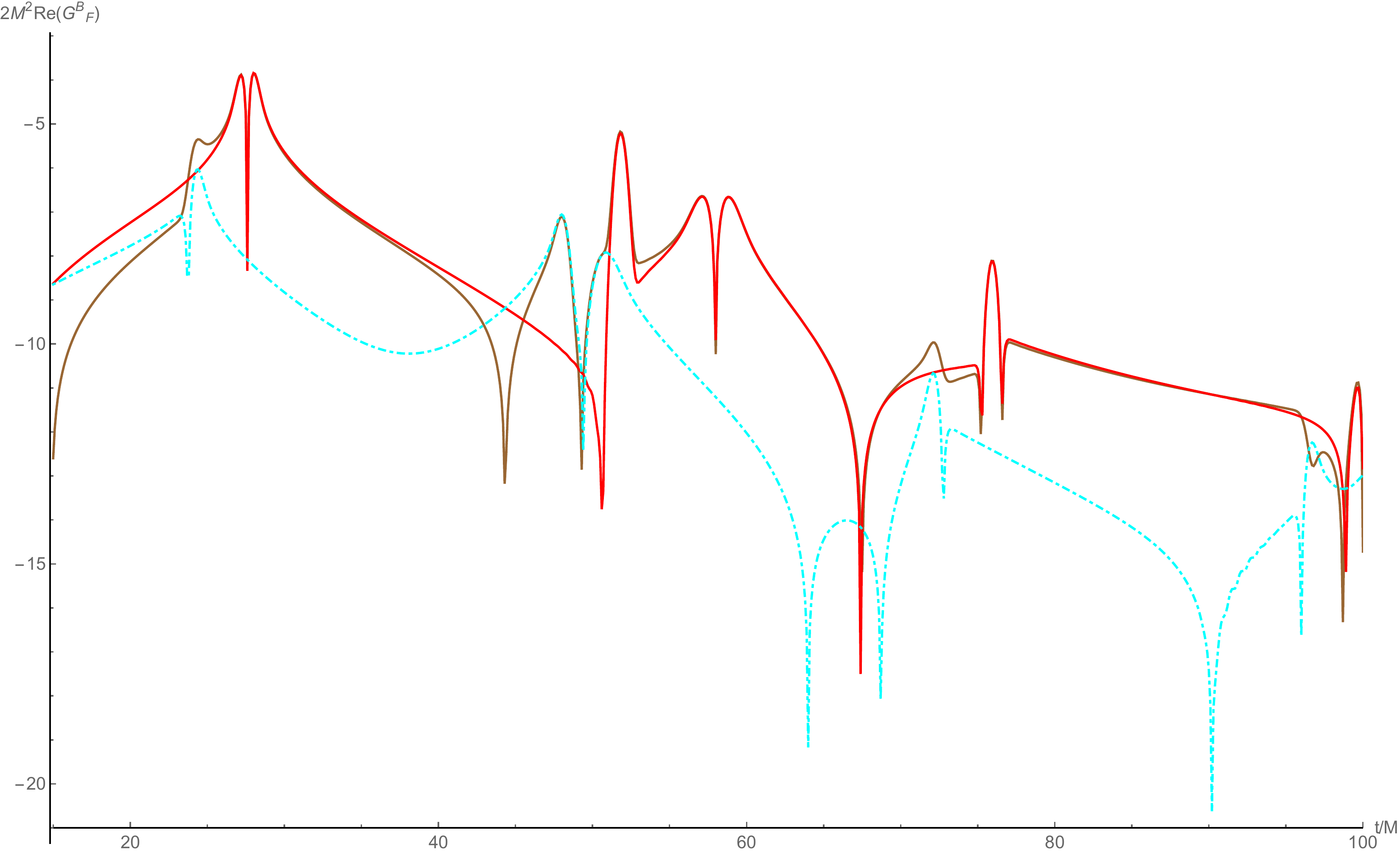}
\caption{
Plots of the RGF and the real part of the FGF (times $2$) for points $x$ and $x'$ on a timelike circular geodesic at $r=6M$.
Top plot (cf.Fig.1~\cite{CDOWa}): RGF calculated with the method of~\cite{CDOWa} (dashed black);
$\text{Re}(G_F^B)$  (the curves of $\text{Re}(G_F^H)$ and $\text{Re}(G_F^U)$ overlap perfectly
with that of $\text{Re}(G_F^B)$ and so we do not include them) calculated with MST (continuous red). 
Bottom plot: log-plot of the (absolute value times $2$ of the) real part of the FGF for the Boulware state (dashed red), the `in' term contribution to it (continuous brown)  
and  the `up' term contribution to it (dot-dashed cyan).
}
\label{fig:ReGFB}
\end{figure}

In Fig.\ref{fig:ReGFB} we plot the real part of the FGF (times $2$) as well as the RGF -- they should agree 
as per Eq.\eqref{eq:RGF via FGF}.
We note some features:
\begin{itemize}
\item The agreement in the top plot between the RGF and the real part of the FGF
is remarkable given that they were calculated using very different methods and with different smoothing functions.
The slight difference in the height of the peaks 
is
 due to using a more severe smoothing in the FGF -- we
checked that increasing $\ell_{cut}$ makes the heights 
coincide with those of RGF but then some spurious oscillations appear, and so we decided to keep
$\ell_{cut}=12$.

\item It displays the known fourfold singularity structure of Eq.\eqref{4-fold} with the singularities `smoothed out'
(the initial $\delta(\sigma)$ singularity is not displayed since the plot is for $t>0$). 

\item Up until shortly before the first light crossing (namely, for $t\lesssim 15M$), our mode-sum calculation --like those in~\cite{CDOWa,Zenginoglu:2012xe,WardellRealFreqRGFSchw,GFKerr}-- does not perform well.
In this `quasilocal' region,  the RGF is calculated via the Hadamard form Eq.\eqref{eq:Hadamard RGF}~\cite{CDOWb,CDOWa,Zenginoglu:2012xe}.
It is not clear how one could calculate the FGF via the Hadamard form Eq.\eqref{eq:Feynman GF Had} since  the biscalar $W$ is in principle not known.
Therefore, we do not plot the FGF in the `quasilocal' region.

\item
The  bottom plot
shows that the dominant contribution to the RGF calculated as the real part of the FGF comes from the `in' modes and that, in particular, the
divergences at light-crossings seem to be due to these modes.

\end{itemize}

\begin{figure}
\begin{tabular}{cc}
\includegraphics[width=8.7cm]{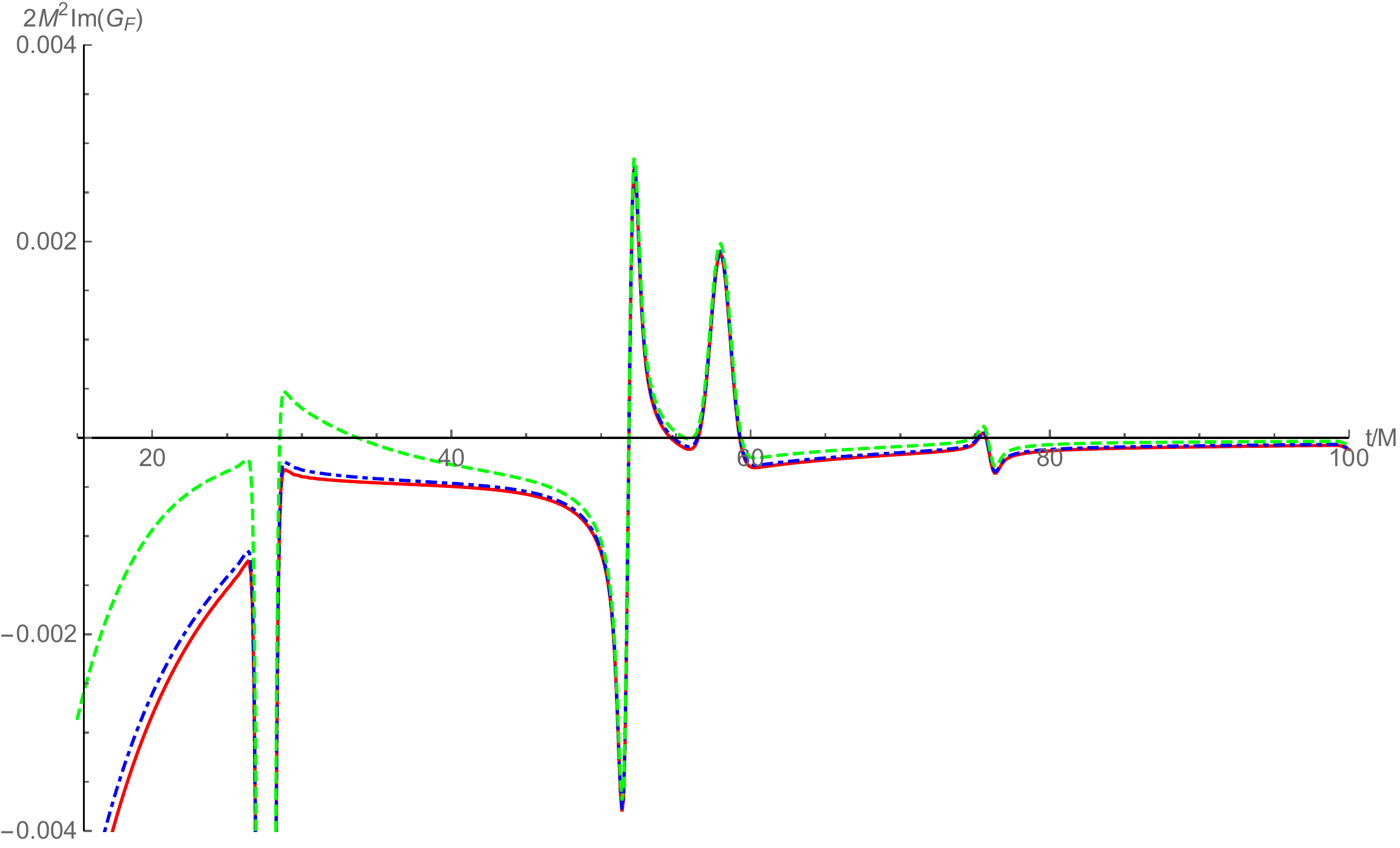}
\end{tabular}
\caption{
Plots of the imaginary part (times $2$)  of the FGF for  points $x$ and $x'$ along a timelike circular geodesic at $r=6M$:
Boulware state (continuous red), the Unruh state (dot-dashed blue) and Hartle-Hawking state (dashed green).
}
\label{fig:NonLogImGFBUHAll}
\end{figure}

\begin{figure}
\includegraphics[width=8.7cm]{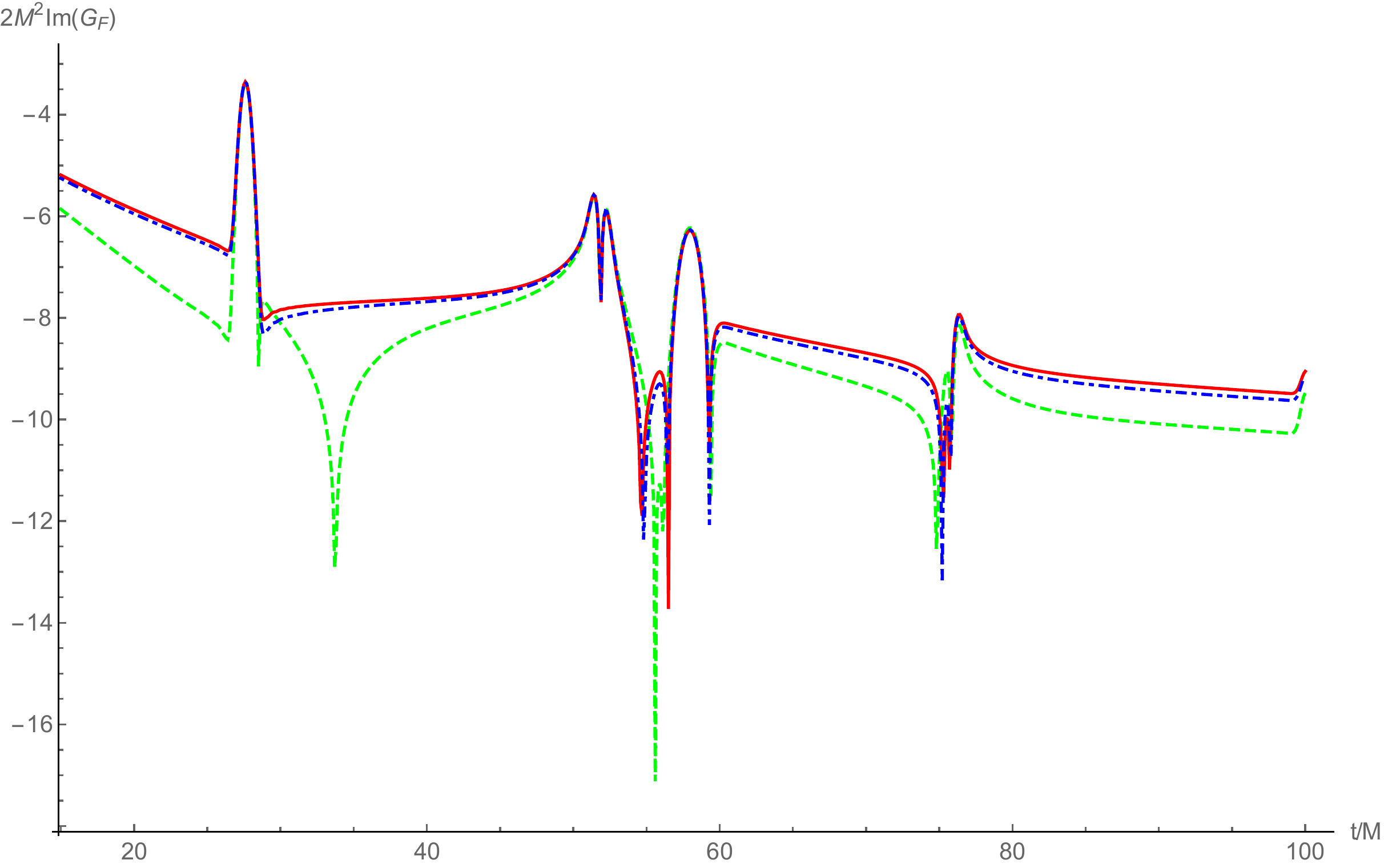}
\\
 \includegraphics[width=8.7cm]{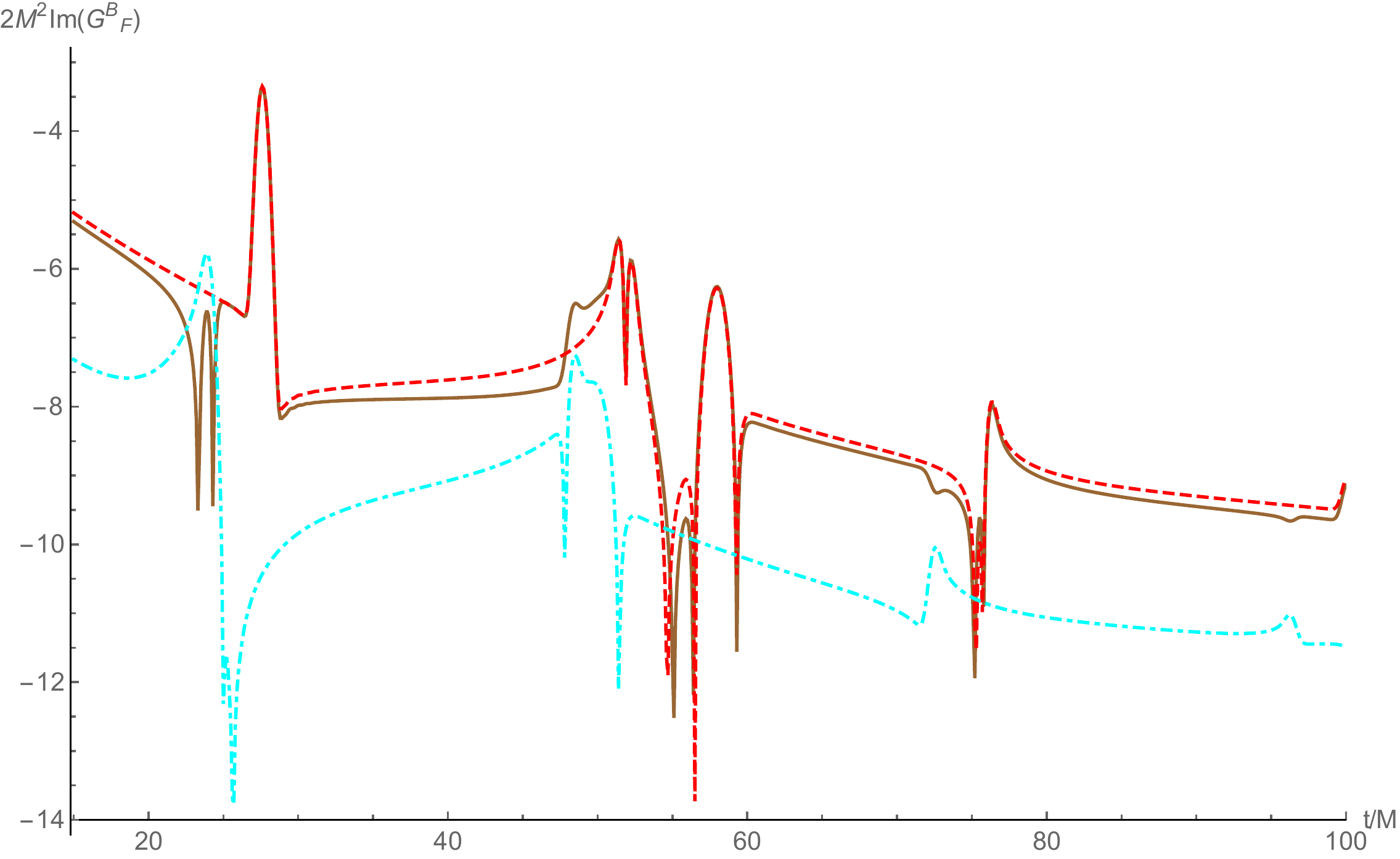}
\\
\includegraphics[width=8.7cm]{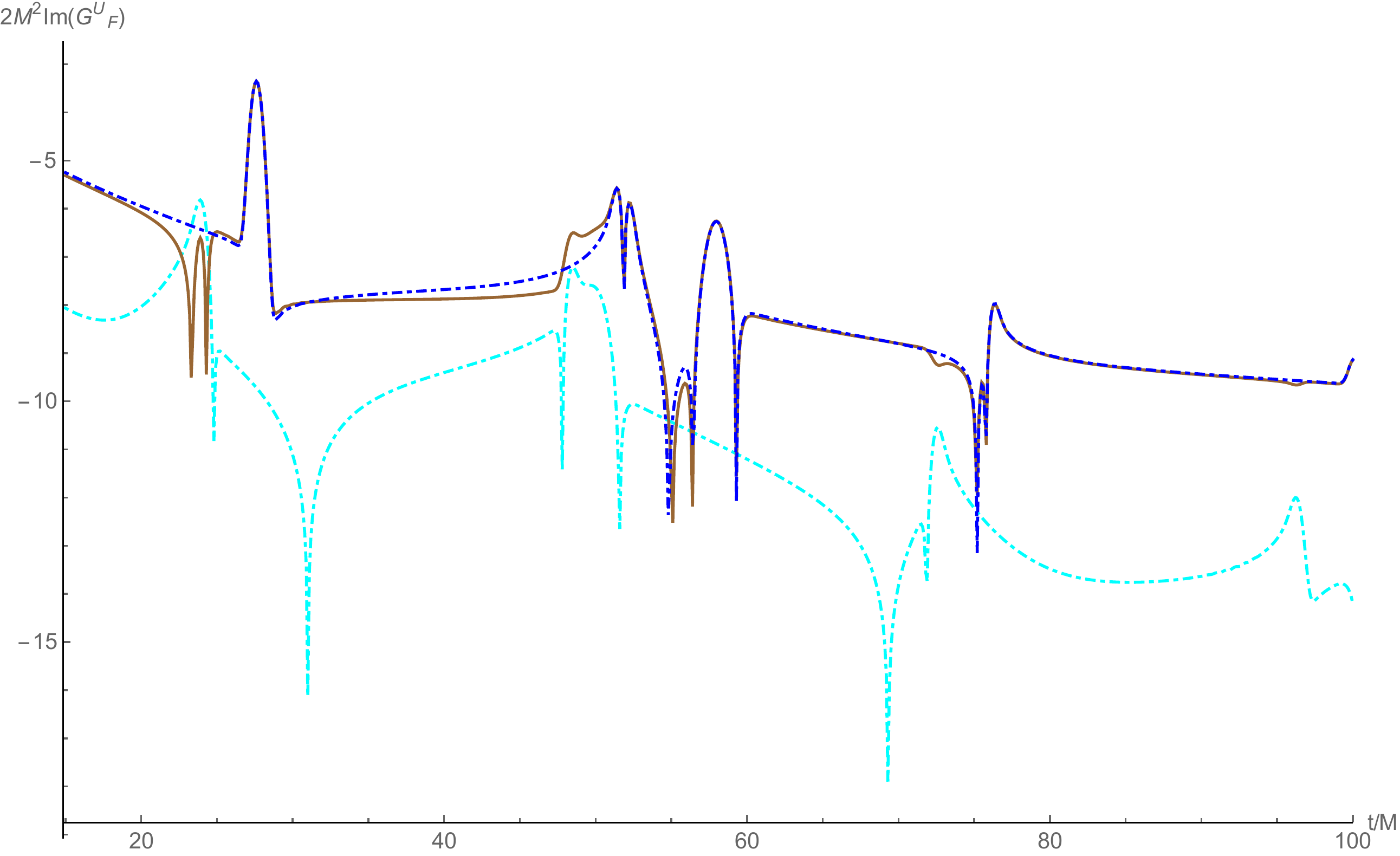}
\\
\includegraphics[width=8.7cm]{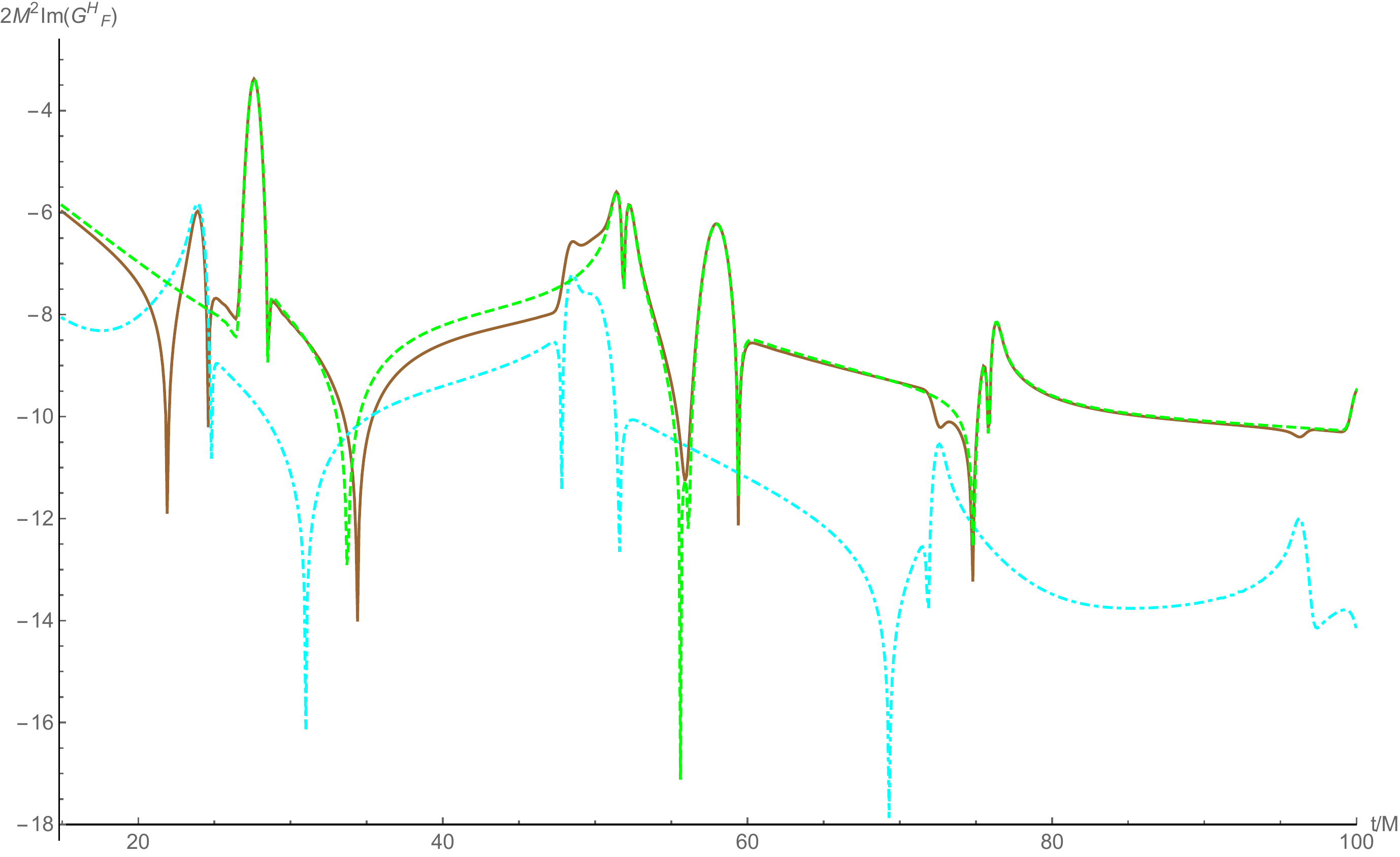}
\caption{
Log-plots of the (absolute value of $2$ times the) imaginary part of the FGF for  points $x$ and $x'$ along a timelike circular geodesic at $r=6M$. 
Top 
plot (this is the log-plot version of Fig.\ref{fig:NonLogImGFBUHAll}): 
FGF for Boulware  (continuous red),  Unruh  (dot-dashed blue) and Hartle-Hawking (dashed green) states;
the differences between the curves correspond to  different correlations in the different quantum states. 
Second plot from the top:
FGF for the Boulware state (dashed red), the `in' term contribution to it (continuous brown)  
and  the `up' term contribution to it (dot-dashed cyan).
Third plot from the top:
FGF for the Unruh state (dashed blue),  the `in' term contribution to it (continuous brown)  
and the `up' term contribution to it (dot-dashed cyan). 
Bottom
plot:  FGF for the Harte-Hawking state (dashed green),  the `in' term contribution to it (continuous brown)  
and  the `up' term contribution to it (dot-dashed cyan). 
}
\label{fig:ImGFBUHAll}
\end{figure}

In 
Figs.\ref{fig:NonLogImGFBUHAll}--\ref{fig:ImGFBUHAll}
we plot the imaginary part of the FGF for the Boulware, Unruh and Hartle-Hawking states.
We note some features:
\begin{itemize}

\item 
The curves 
for FGF for all states
 display the conjectured fourfold singularity structure of Eq.\eqref{4-fold FGF}. 


\item 
The quantum correlations are dominant for points which are joined by a null geodesic, as expected (and as it already
happens in flat space-time).
The form and location of the divergences at these light-crossings are state-independent and
so they `harness' the form of the correlator to some extent.

\item 
The visible differences between the three curves 
in 
Fig.\ref{fig:NonLogImGFBUHAll} and
the top 
 plot of Fig.\ref{fig:ImGFBUHAll}
are physical differences due to  
different correlations in the three different quantum states.

\item
The  
three
 bottom plots of Fig.\ref{fig:ImGFBUHAll}
show that the dominant contribution to the FGF for all three states comes from the `in' modes, similarly to the RGF above.

\mc{Can we say anything more about the quantum correlations...?}

\end{itemize}

Heuristically, the reason why the `in' modes contribution to both RGF and FGF dominates over the `up' modes contribution is probably
 the following. The divergences at light-crossings arise from the large-$\ell$ modes in the $\ell$-sum.
Now, for large-$\ell$, the radial potential in Eq.\eqref{eq:potential} is highly-peaked at a radius $r_0$ near the unstable photon orbit, which is located at $r=3M$. Therefore, the `in' [resp. `up'] modes are mostly `trapped' to its right, 
$r>r_0\approx 3M$ [resp. left,  $r<r_0\approx 3M$].
It is thus reasonable that the `in' modes are the dominant contribution to the RGF at $r=6M>r_0$.

\section{Final Comments}\label{sec:end}

We have calculated, for the first time in the literature, the quantum correlator for a scalar field outside a Schwarzschild black hole.
The explicit calculation of the correlator manifests the global fourfold singularity structure which has been
previously
 conjectured
 and
shows the different correlations for the different quantum states.
In the future it will be interesting to extend this work to calculate the correlator with one point inside the horizon and one point outside, in order to observe the correlations
between in-falling and outgoing Hawking particles.

\section*{Acknowledgments}
C.B. acknowledges the financial support received from the National Council for Scientific and Technological Development, CNPq  (Brazil), through a M.Sc.~scholarship.
M.C. is thankful to Barry Wardell, Adrian Ottewill and William G.~Unruh for useful discussions. 
M.C. thanks the University of British Columbia, Canada, for hospitality while this work was in progress. 
M.C. acknowledges partial financial support by CNPq (Brazil), process number 308556/2014-3. 
This work makes use of the Black Hole Perturbation Toolkit~\cite{BHPT} (the MST code part of it will be added in the near future).


\bibliographystyle{apsrev}

\end{document}